\documentclass[10pt,twocolumn,english,superscriptaddress,floatfix,tightenlines,aps,prl]{revtex4-1}
\usepackage[T1]{fontenc}
\usepackage[utf8]{inputenc}
\usepackage{geometry}
\usepackage{xcolor}
\usepackage{pdfcolmk}
\usepackage{babel}
\usepackage{units}
\usepackage{amsmath}
\usepackage{amssymb}
\usepackage{graphicx}
\PassOptionsToPackage{normalem}{ulem}
\usepackage{ulem}
\usepackage[unicode=true,pdfusetitle,
 bookmarks=true,bookmarksnumbered=false,bookmarksopen=false,
 breaklinks=false,pdfborder={0 0 1},backref=false,colorlinks=false]
 {hyperref}

\makeatletter

\providecolor{lyxadded}{rgb}{0,0,1}
\providecolor{lyxdeleted}{rgb}{1,0,0}
\DeclareRobustCommand{\lyxsout}[1]{\ifx\\#1\else\sout{#1}\fi}

\usepackage{babel}
\usepackage{tikz}
\usetikzlibrary{decorations.pathreplacing,calc}
\usepackage{lipsum}
\usepackage{filecontents}
\RequirePackage{atbegshi}
\makeatother

\begin{document}

\title{Collapse of critical nematic fluctuations in FeSe under pressure}

\author{Pierre Massat}
\email{pierre.massat@paris7.jussieu.fr}

\affiliation{Laboratoire Matériaux et Phénomènes Quantiques, UMR 7162 CNRS, Université
Paris Diderot, Paris, France}

\author{Yundi Quan}

\affiliation{Department of Physics and Center for Advanced Quantum Studies, Beijing
Normal University, Beijing, 100875 China}

\author{Romain Grasset}

\affiliation{Laboratoire Matériaux et Phénomènes Quantiques, UMR 7162 CNRS, Université
Paris Diderot, Paris, France}

\author{Marie-Aude Méasson}

\affiliation{Laboratoire Matériaux et Phénomènes Quantiques, UMR 7162 CNRS, Université
Paris Diderot, Paris, France}

\affiliation{Université Grenoble Alpes, CNRS, Grenoble INP, Institut Néel, F-38000
Grenoble, France}

\author{Maximilien Cazayous}

\author{Alain Sacuto}

\affiliation{Laboratoire Matériaux et Phénomènes Quantiques, UMR 7162 CNRS, Université
Paris Diderot, Paris, France}

\author{Sandra Karlsson}

\thanks{Present address: Department of Quantum Matter, Physics University
of Geneva, 24 Quai Ernest-Ansermet - 1211 Geneva 4}

\author{Pierre Strobel}

\author{Pierre Toulemonde}

\affiliation{Université Grenoble Alpes, CNRS, Grenoble INP, Institut Néel, F-38000
Grenoble, France}

\author{Zhiping Yin}
\email{yinzhiping@bnu.edu.cn}

\affiliation{Department of Physics and Center for Advanced Quantum Studies, Beijing
Normal University, Beijing, 100875 China}

\author{Yann Gallais}
\email{yann.gallais@paris7.jussieu.fr}

\affiliation{Laboratoire Matériaux et Phénomènes Quantiques, UMR 7162 CNRS, Université
Paris Diderot, Paris, France}

\date{\today}
\begin{abstract}
We report the evolution of the electronic nematic susceptibility in
FeSe via Raman scattering as a function of hydrostatic pressure up
to $\unit[5.8]{GPa}$ where the superconducting transition temperature
$T_{c}$ reaches its maximum. The critical nematic fluctuations observed
at low pressure vanish above $\unit[1.6]{GPa}$, indicating
they play a marginal role in the four-fold enhancement of $T_{c}$
at higher pressures. The collapse of nematic fluctuations appears
to be linked to a suppression of low energy electronic excitations
which manifests itself by optical phonon anomalies at around $\unit[2]{GPa}$,
in agreement with lattice dynamical and electronic structure calculations
using local density approximation combined with dynamical mean field
theory. Our results reveal two different regimes of nematicity in
the phase diagram of FeSe under pressure: a $d$-wave Pomeranchuk
instability of the Fermi surface at low pressure and a magnetic driven
orthorhombic distortion at higher pressure.
\end{abstract}
\maketitle
%\begin{filecontents}{additional_refs.bib}
%@Misc{QuanTBP,
%author = {Quan, Y. and Yin, Z. P.},
%howpublished = {To be published}}
%@Misc{Massat2017SM,
%howpublished = {See Supplementary Material where references \citep{Boehmer2013,Karlsson2015,Gnezdilov2013,Allen1972,Wien2k,Haule2010,Haule2015,Haule2016,Haule2007,Werner2006,Kumar2010a,QuanTBP,Yin2013,Yin2011,Yin2011a,Abrikosov1975,Shimojima2014,Fanfarillo2016,Kushnirenko2017,Guterding2017,Boeri2002,Goncharov2003a} are cited and for more detailed experimental and theoretical results and analysis.}}
%\end{filecontents}

The pairing mechanism of iron-based superconductors (Fe SC) is believed
to result from interband spin fluctuations \citep{Kuroki2008,Mazin2008}.
The spin fluctuations scenario is motivated by the observation that
the maximum superconducting transition temperature corresponds to
the end point of a stripe-like magnetic phase in several Fe SC \citep{Johnston2010}.
However the magnetic order is invariably preceded by, or concomitant
with, an electron nematic phase whereby the electronic sub-system
spontaneously breaks the four-fold tetragonal symmetry and induces
an orthorhombic distortion of the lattice \citep{Chu2010,Fernandes2014-01}.
In several Fe SC strong electron nematic fluctuations (NF) have been
observed near the optimal critical temperature $T_{c}$ \citep{Chu2012,Gallais2013,Kuo2016},
hinting that they could play a role too in the pairing mechanism.
Recent theoretical works indeed support the idea that critical NF
near a quantum critical point (QCP) are generically helpful for SC
pairing \citep{Maier2014,Lederer2015,Labat2017}. However addressing
the driving force behind electron nematicity and the role of critical
NF in enhancing $T_{c}$ remains a challenge in many Fe SC, as magnetic
and nematic orderings often occur simultaneously.

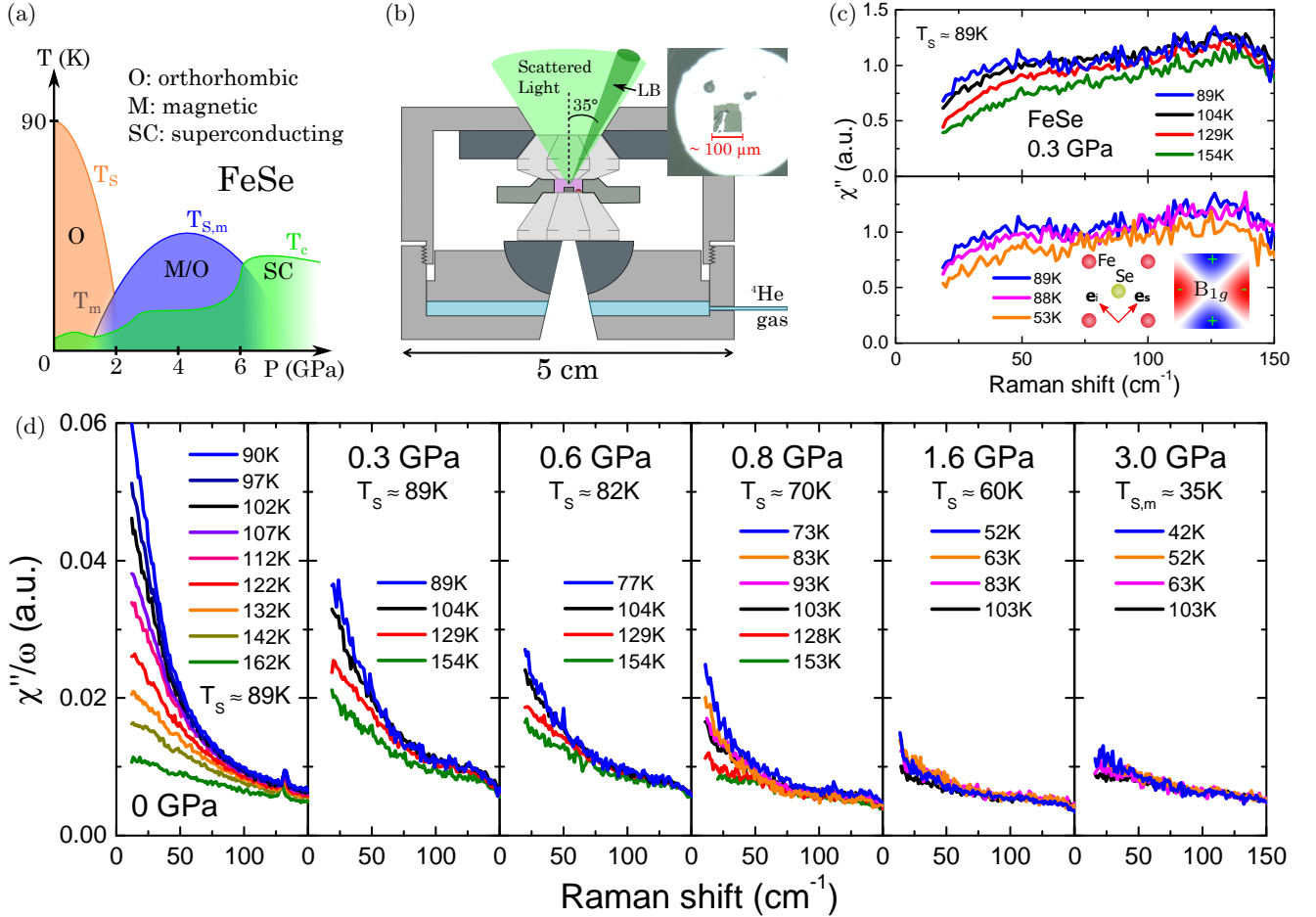
\begin{figure*}[t]
\begin{tikzpicture}[baseline={([yshift={-\ht\strutbox}]current bounding box.north)}]%the baseline=... option allows to align all elements vertically to the top
\node[anchor=south west,inner sep=0] (pc) at (0,0) {\includegraphics[height=4.5cm]{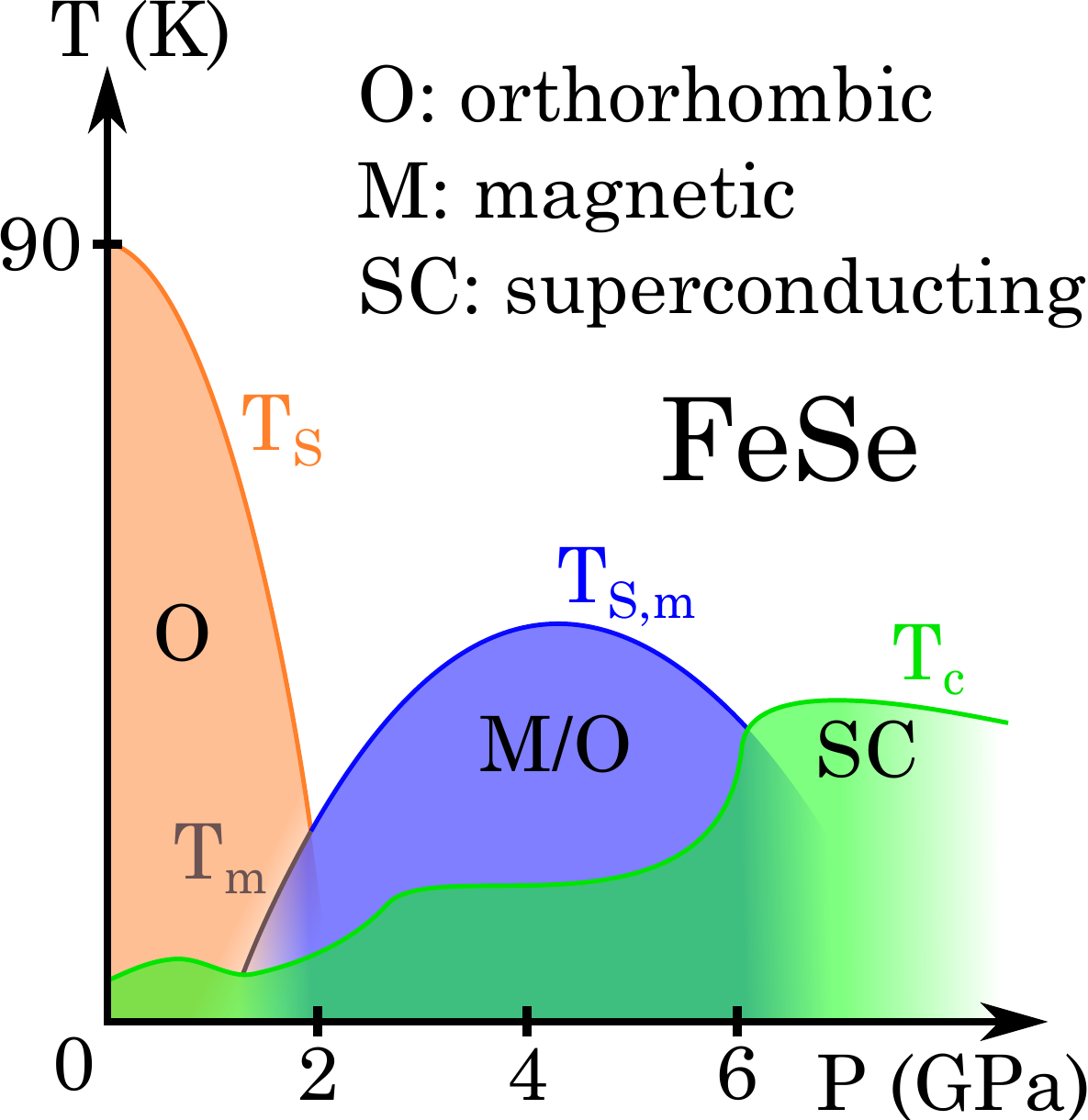}

};
\begin{scope}[x={(pc.south east)},y={(pc.north west)}]
\node at (0.,1.1) {(a)};
\end{scope}
\end{tikzpicture}\hfill{}\begin{tikzpicture}[baseline={([yshift={-\ht\strutbox}]current bounding box.north)}]%the baseline=... option allows to align all elements vertically to the top
\node[anchor=south west,inner sep=0] (pc) at (0,0) {\includegraphics[height=4.5cm]{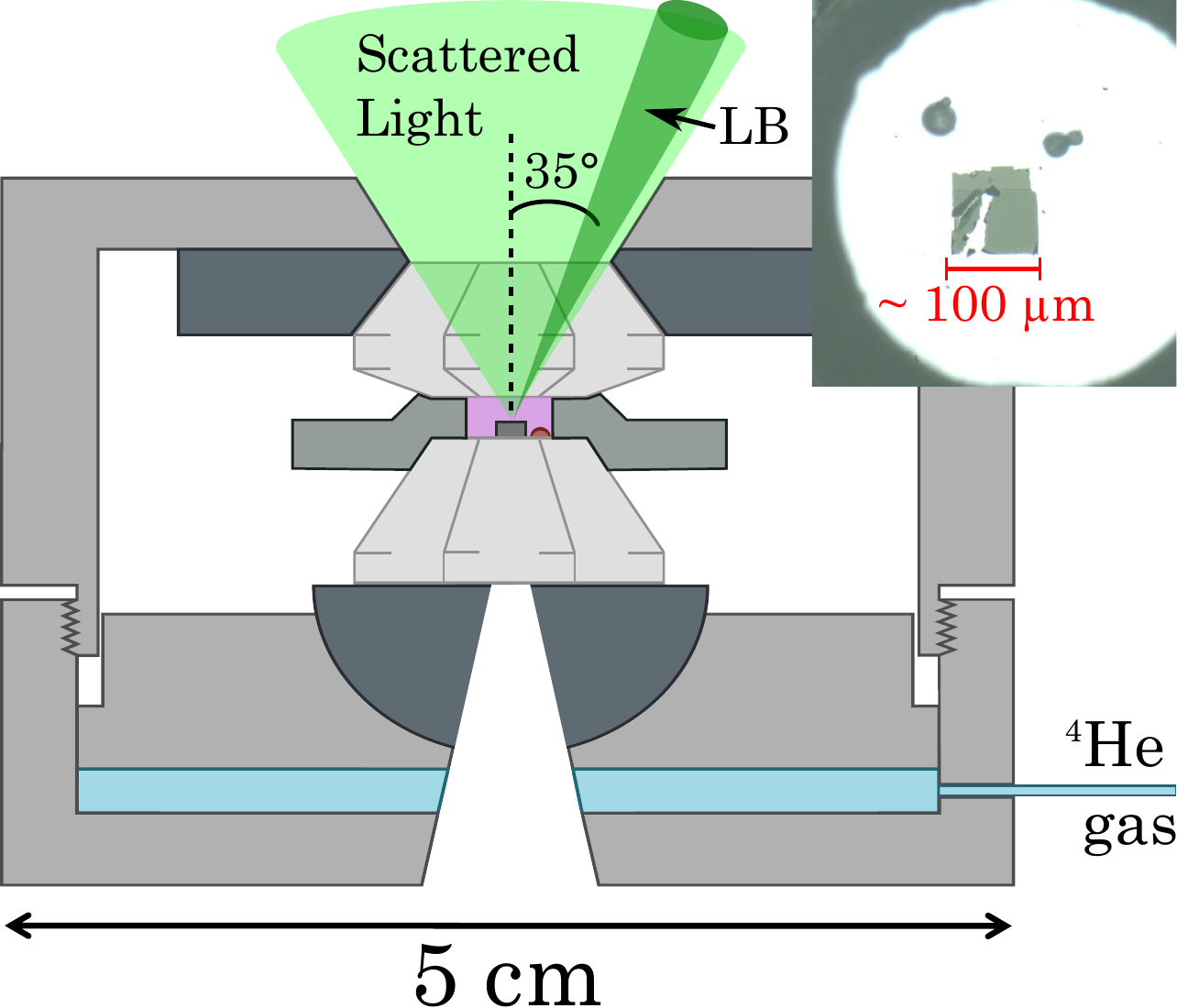}

};
\begin{scope}[x={(pc.south east)},y={(pc.north west)}]
\node at (0.,1.1) {(b)};
\end{scope}
\end{tikzpicture}\hfill{}\begin{tikzpicture}[baseline={([yshift={-\ht\strutbox}]current bounding box.north)}]%the baseline=... option allows to align all elements vertically to the top
\node[anchor=south west,inner sep=0] (rr) at (0,0) {\includegraphics[height=5.5cm]{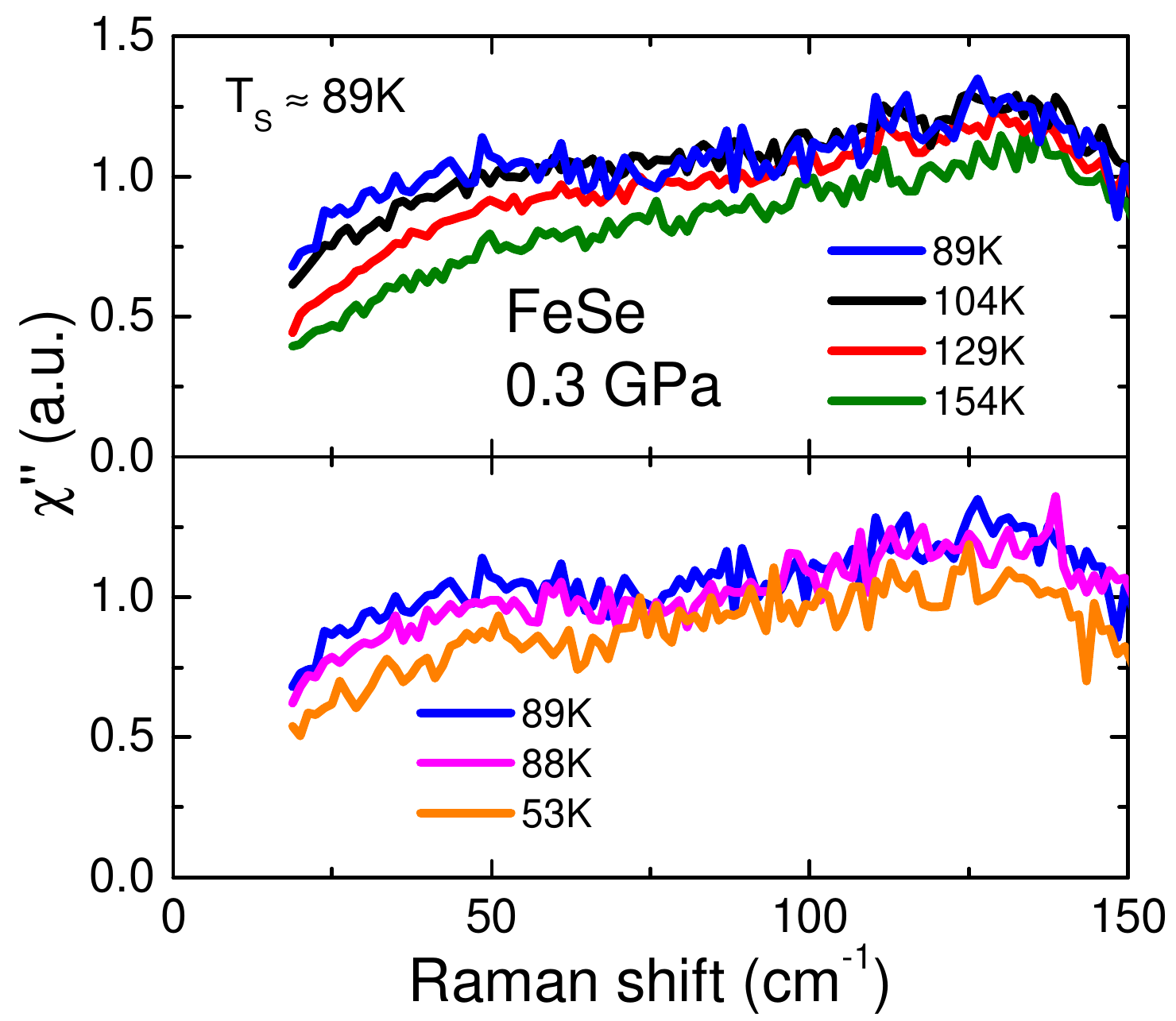}

};
\begin{scope}[x={(rr.south east)},y={(rr.north west)}]
\node at (.035,.95) {(c)};
\node (vertex) at (0.625,0.27) {

\includegraphics[height=1cm]{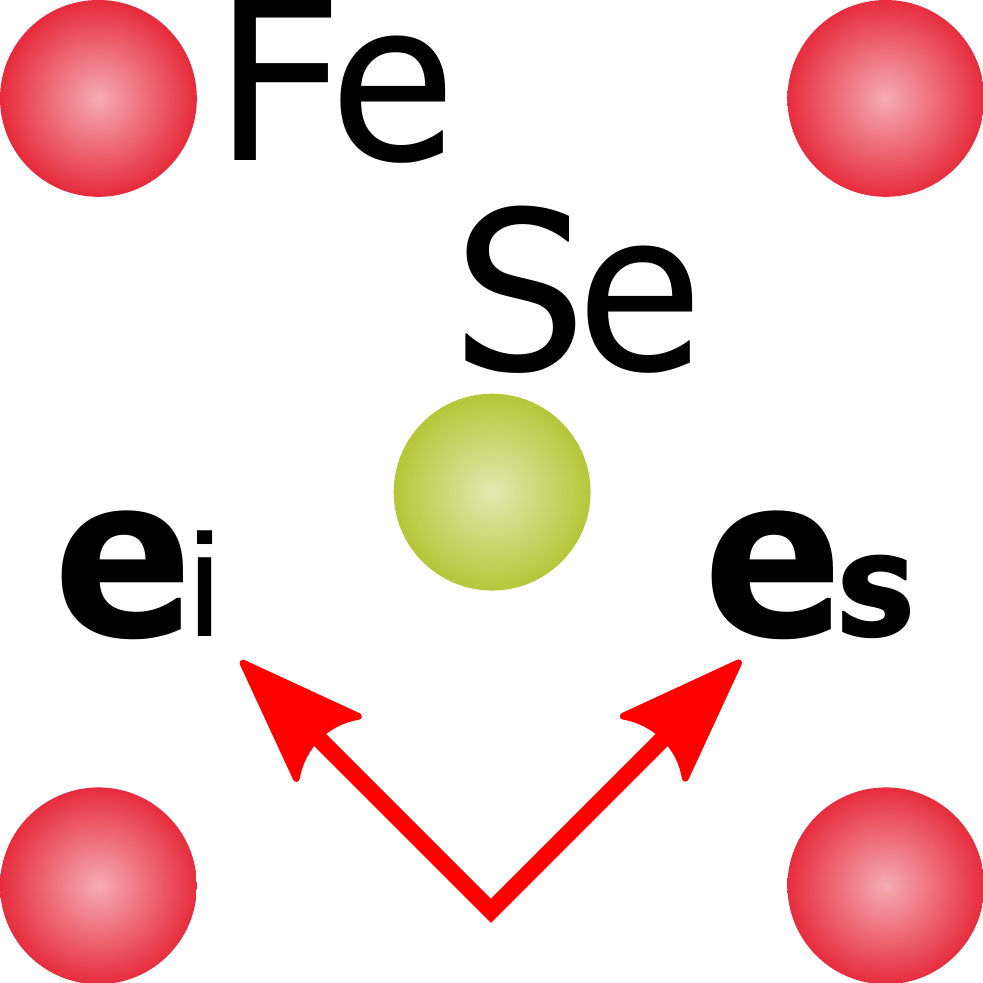}

};
\node (vertex) at (0.825,0.27) {\includegraphics[width=0.12\columnwidth]{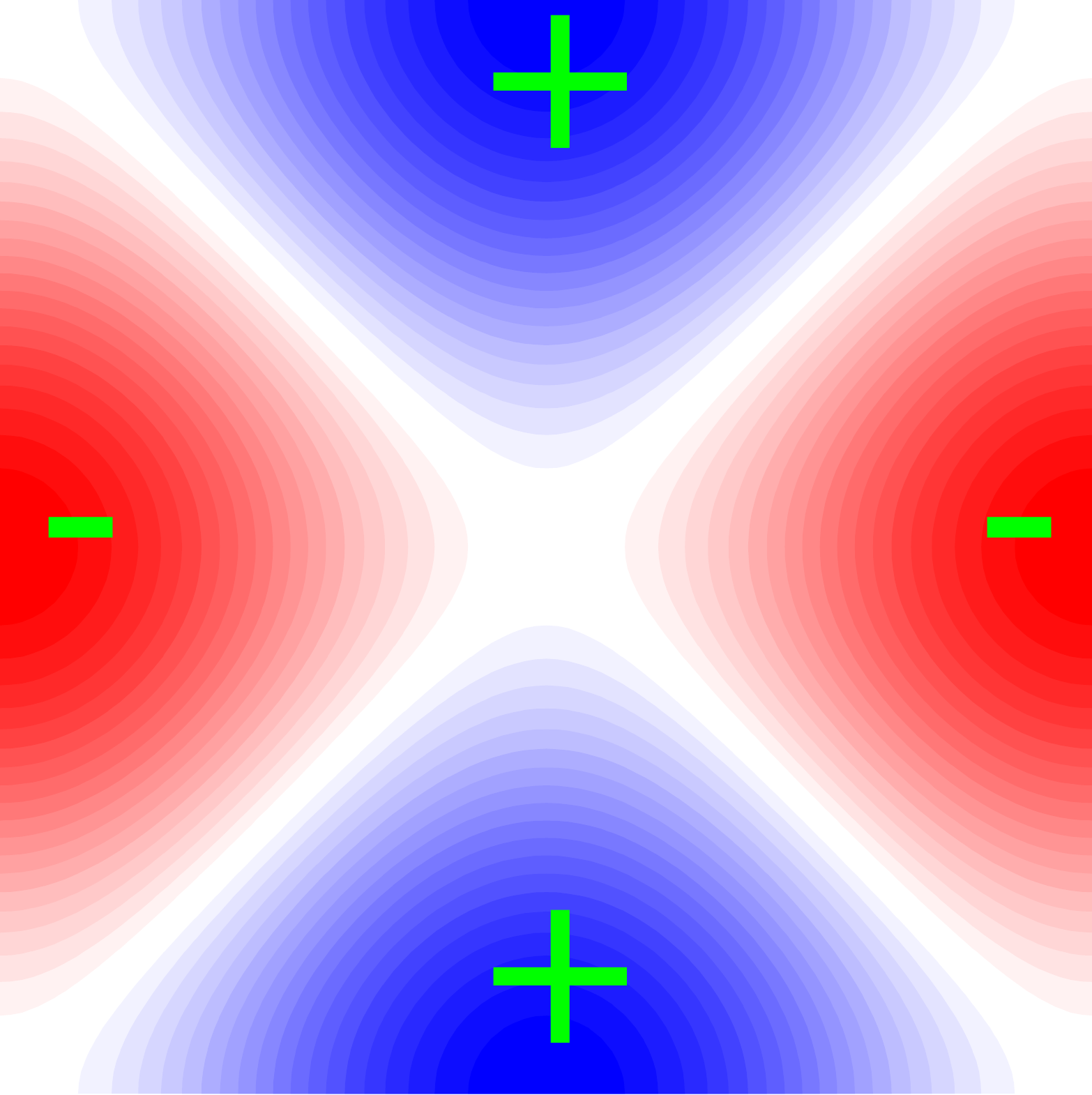}

};
\node at (vertex.center) {\footnotesize B$_{1g}$};
\end{scope}
\end{tikzpicture}

\begin{tikzpicture}
\node[anchor=south west,inner sep=0] (rc) at (0,0) {\includegraphics[width=1\textwidth]{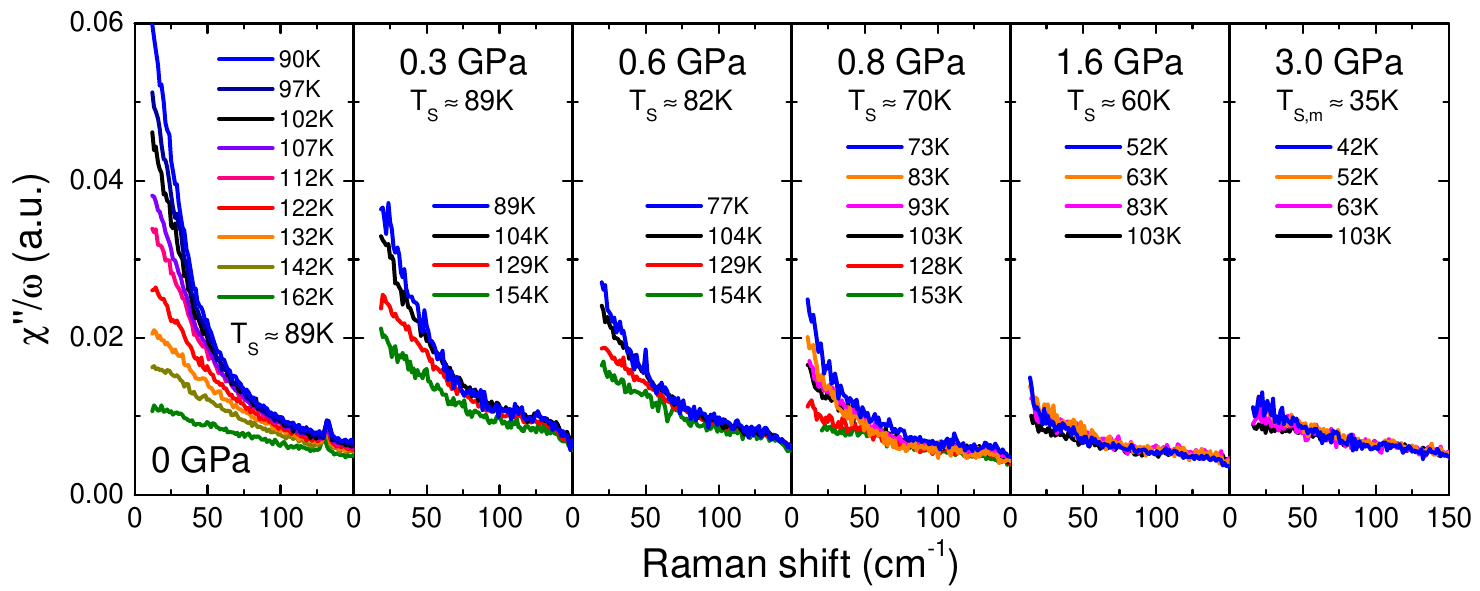}

};
\begin{scope}[x={(rc.south east)},y={(rc.north west)}]
\node at (0.025,.95) {(d)};
\end{scope}
\end{tikzpicture}

\caption{Raman spectroscopy of FeSe under hydrostatic pressure. (a) Schematic
phase diagram of FeSe under pressure. (b) Side view sketch of the
Raman pressure cell. LB stands for Laser Beam. Inset shows a top view
photograph of the sample inside the pressure cell, with three ruby
balls above it. (c) Temperature dependence above (upper panel) and
below $T_{S}$ (lower panel) of the $B_{1g}$ Raman response at $\unit[0.3]{GPa}$.
Insets show the $x'y'$ polarizations configuration and the amplitude
of the corresponding $B_{1g}$ Raman vertex in momentum space. (d)
Temperature dependent Raman conductivity $\chi''\left(\omega\right)/\omega$
measured at $P=$ 0, 0.3, 0.6, 0.8, 1.6 and $\unit[3.0]{GPa}$. The
data at $\unit[0.3]{GPa}$ are the same as that of (c) (divided by
frequency).\label{fig:Raman-FeSe_pressure}}
\end{figure*}

In this context superconducting FeSe stands out for its unusual properties
compared to other Fe SC \citep{Coldea2018}. In bulk form and at ambient
pressure, it has relatively low $T_{c}$ but displays a nematic phase
without any sign of magnetic ordering, thus challenging magnetic driven
scenarios of nematicity \citep{Hsu2008,McQueen2009a,Baek2014,Boehmer2015,Watson2015}.
A strong increase of $T_{c}$ is observed upon applying hydrostatic
pressure, reaching $\sim\unit[37]{K}$ at $\sim\unit[6]{GPa}$ \citep{Medvedev2009,Garbarino2009}.
While recent ARPES and transport measurements have suggested a possible
link between the increases of $T_{c}$ and changes in Fermi surface
topology in FeSe \citep{Ye2015,Miyata2015,Lei2016,Shiogai2017,Sun2017a},
its pressure phase diagram also differs significantly from other prototypical
Fe SC \citep{Bendele2012,Terashima2015,Sun2016a,Kothapalli2016a,Svitlyk2017,Boehmer2018}.
The nematic phase transition temperature $T_{S}$ initially decreases
with pressure, and merges at around $\unit[1.5-2]{GPa}$ with a pressure
induced magnetic phase which is likely similar to the stripe-ordered
phase observed in other Fe SC \citep{Kothapalli2016a,Wang2016b,Boehmer2018a,Khasanov2018}.
The temperature $T_{S,m}$ of the resulting coupled magneto-structural
transition has a non-monotonic pressure dependence terminating close
to optimal pressure where $T_{c}$ is maximum \citep{Sun2016a} (figure
\ref{fig:Raman-FeSe_pressure}(a)), suggesting the presence of a QCP
and questioning the respective roles of critical magnetic and nematic
fluctuations in the pressure-induced four-fold enhancement of $T_{c}$.

In this letter we report the temperature and pressure dependence of
the NF in bulk FeSe single crystals using Raman spectroscopy up to
$\unit[5.8]{GPa}$. We show that temperature dependent critical NF
disappear rapidly upon increasing pressure and essentially vanish
above $\unit[1.6]{GPa}$, indicating two different regimes of nematicity:
a low pressure regime where the nematic transition is driven by a
$d$-wave Pomeranchuk instability of the Fermi surface, and a higher
pressure regime where it is only a secondary symmetry breaking induced
by the magnetic transition, implying that the enhancement of $T_{c}$
above $\unit[2]{GPa}$ is not driven by critical NF in this pressure
range. We further show that the disappearance of the critical NF is
accompanied by anomalies in the pressure dependence of Raman-active
optical phonons' frequencies. Supported by density functional theory
plus dynamical mean-field theory calculations, we link the phonon
anomalies and the collapse of NF at $\sim\unit[2]{GPa}$ to a Lifshitz
transition of the Fermi surface.

Raman measurements under pressure were performed using a membrane
diamond anvil cell (DAC) allowing continuous change of pressure at
low temperature, and designed with a large numerical aperture as described
in \cite{Buhot2015,Grasset2018}. Helium was used as the pressure
transmitting medium. Figure \ref{fig:Raman-FeSe_pressure}(b) shows
a sketch of the pressure cell and a photograph of the sample inside
the cell. Focus was made on Raman spectra taken in the $B_{1g}$ symmetry
\citep{Massat2017SM} which, in the 1-Fe unit cell notation, corresponds
to the nematic order observed in FeSe at ambient pressure \citep{Massat2016}.

Figure \ref{fig:Raman-FeSe_pressure}(c) displays the temperature
dependence of the Raman response in $B_{1g}$ symmetry at $\unit[0.3]{GPa}$.
It follows the behavior of the Raman spectra of FeSe at ambient pressure
reported earlier \citep{Massat2016}. When approaching the nematic
transition from high temperatures, the growth of critical NF manifests
itself by an increase in the low energy Raman response $\chi''\left(\omega\right)$
upon approaching $T_{S}$, and a subsequent decrease in the nematic
phase below $T_{S}$.

The temperature dependence of the $B_{1g}$ Raman conductivity $\chi''/\omega$,
which controls the static nematic susceptibility (see equation \ref{eq:chi0}
below), is plotted at different pressures in figure \ref{fig:Raman-FeSe_pressure}(d).
For clarity only spectra above the estimated $T_{S}$, which is $\unit[89]{K}$
at $\unit[0]{GPa}$ but decreases when increasing pressure \citep{Miyoshi2013,Terashima2015},
are shown. The spectra at 0 GPa were taken outside the pressure cell
on a different crystal from the same batch, and are shown here for
comparison. In the Raman conductivity spectrum the NF appear as a
quasi-elastic peak (QEP) centered at zero-energy \cite{Gallais2016}.
At low pressures, below $\unit[0.8]{GPa}$, the QEP intensity increases
significantly when lowering temperature down to $T_{S}$, following
the behavior observed at ambient pressure. However upon increasing
pressure the maximum intensity of the QEP close to $T_{S}$ (blue
curve at each pressure), and its overall enhancement decrease significantly.
At $\unit[1.6]{GPa}$ and $\unit[3.0]{GPa}$ the QEP is barely visible
and shows negligible enhancement upon cooling.

\begin{figure}
\begin{tikzpicture}
\node[anchor=south west,inner sep=0] (pc) at (0,0) {\includegraphics[width=1\columnwidth]{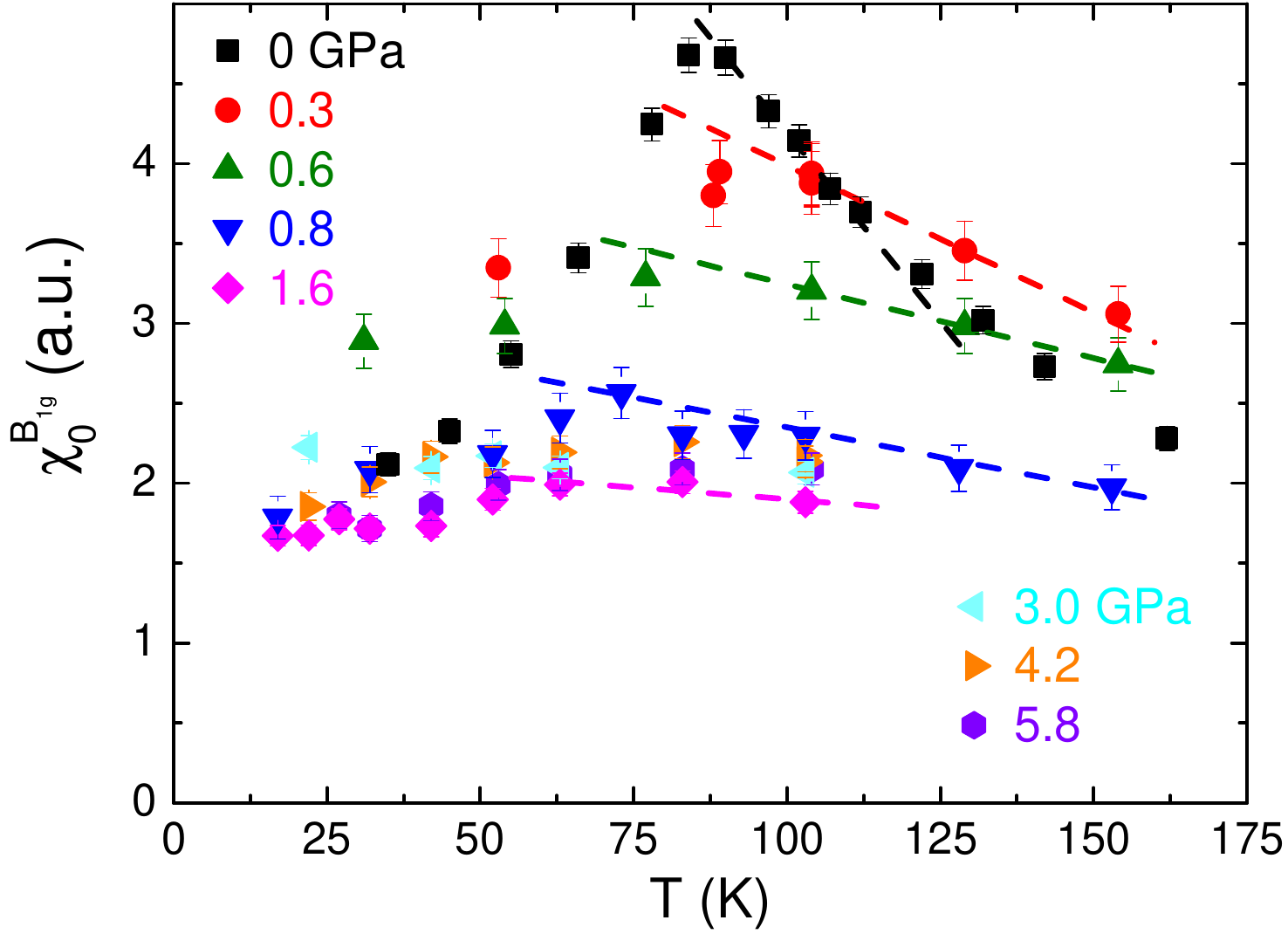}

};
\begin{scope}[x={(pc.south east)},y={(pc.north west)}]
\node at (0.04,.95) {(a)};
\end{scope}
\end{tikzpicture}

\smallskip{}
\begin{tikzpicture}
\node[anchor=south west,inner sep=0] (pc) at (0,0) {\includegraphics[width=1\columnwidth]{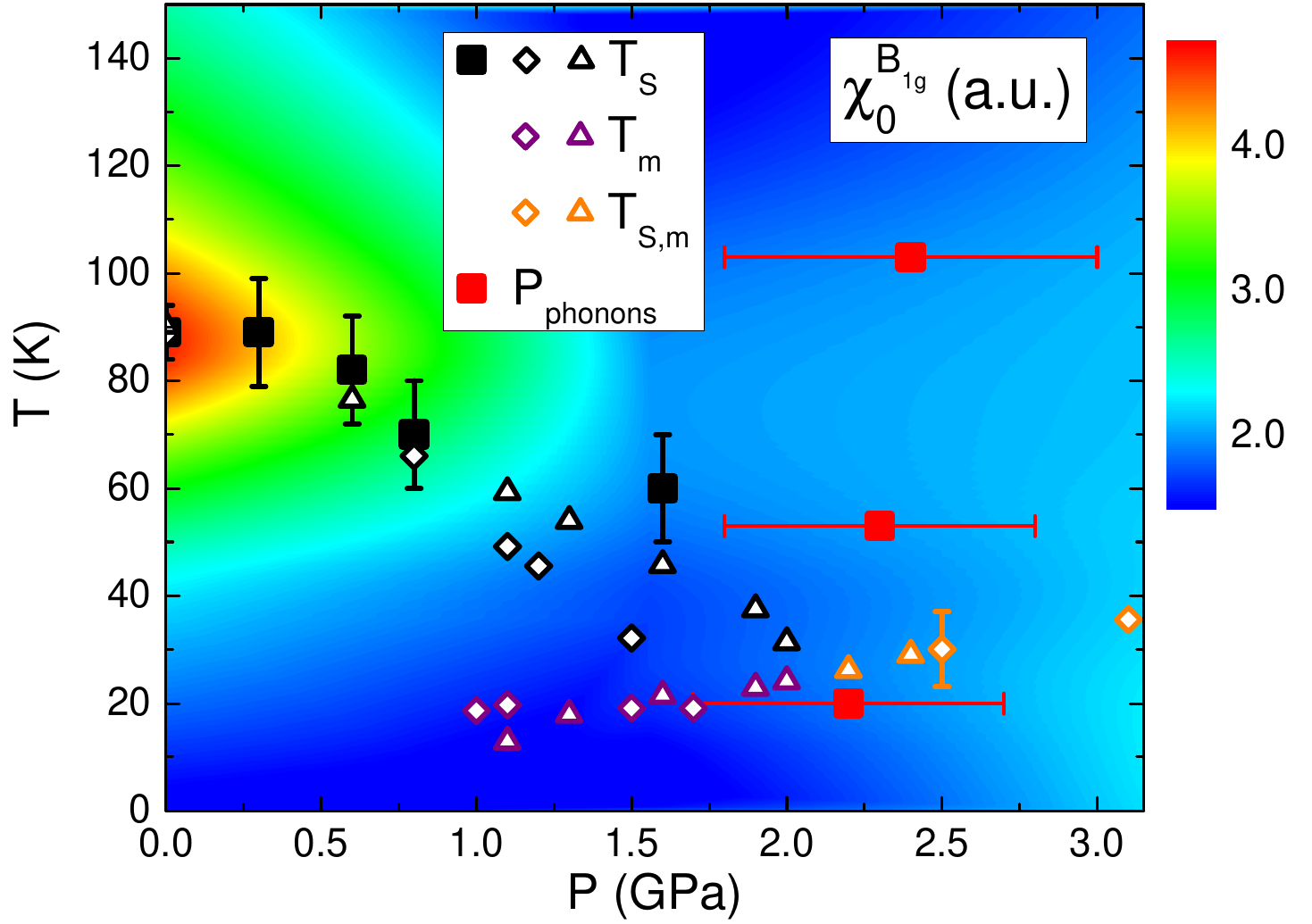}

};
\begin{scope}[x={(pc.south east)},y={(pc.north west)}]
\node at (0.03,.95) {(b)};
\end{scope}
\end{tikzpicture}
\raggedright{}\caption{(a) Temperature dependence of the static nematic susceptibility $\chi_{0}^{B_{1g}}$
at different pressures between 0 and $\unit[5.8]{GPa}$. Dashed lines
are linear fits showing the slope at $\unit[100]{K}$: $\left.\frac{\mathrm{d}\chi_{0}^{B_{1g}}}{\mathrm{d}T}\right|_{\unit[100]{K}}$
at each pressure from 0 to $\unit[1.6]{GPa}$; within experimental
accuracy, the slope is 0 above $\unit[3]{GPa}$. (b) Color map of
the static nematic susceptibility $\chi_{0}^{B_{1g}}$ as a function
of temperature and pressure, plotted using the data points in (a).
The color scale is in arbitrary units. The temperatures of the structural
($T_{S}$), magnetic ($T_{m}$) and magneto-structural ($T_{S,m}$)
transitions are indicated by black, purple and orange symbols respectively.
Full squares: this work; empty diamonds: \cite{Kothapalli2016a};
empty triangles: \cite{Wang2016b}. Red squares with error bars indicate
the pressure range of the phonon anomalies at $\unit[103]{K}$, $\unit[53]{K}$
and $\unit[20]{K}$ (see text).\label{fig:chi0}}
\end{figure}

In order to quantify the observed pressure evolution of NF, we computed
the static $B_{1g}$ nematic susceptibility $\chi_{0}^{B_{1g}}$,
obtained from the Raman conductivity through the Kramers-Kronig relation:

\begin{equation}
\chi_{0}\left(T,P\right)=\int_{0}^{\infty}\frac{\chi''\left(\omega,T,P\right)}{\omega}\mathrm{d}\omega\label{eq:chi0}
\end{equation}
The integral was performed over the whole frequency range accessible
from our data, i.e. from 0 to $\unit[470]{cm^{-1}}$, with the data
at low energy resulting from linear extrapolation of the Raman response
between 0 and $\unit[12]{cm^{-1}}$. The resulting temperature dependence
of the nematic susceptibility at each pressure is plotted in figure
\ref{fig:chi0}(a) where we have included additional points at higher
pressures: $\unit[4.2]{GPa}$ and $\unit[5.8]{GPa}$. At low pressures
$\chi_{0}^{B_{1g}}$ increases when lowering temperature, reaches
a maximum at $T_{S}\left(P\right)$ and decreases inside the nematic
phase. The pressure evolution of $\chi_{0}^{B_{1g}}\left(T\right)$
indicates a change of regime of NF upon increasing pressure: while
the increase of $\chi_{0}^{B_{1g}}$ when lowering temperature down
to $T_{S}$ is still significant at $\unit[0.8]{GPa}$, indicating
sizable critical NF close to $T_{S}$ at this pressure, it is much
weaker at $\unit[1.6]{GPa}$ and there is essentially no increase
at $\unit[3]{GPa}$ and above. Besides, at $\unit[4.2]{GPa}$ and
$\unit[5.8]{GPa}$ the static susceptibility shows a small but clear
suppression below $\sim\unit[40]{K}$ and $\sim\unit[60]{K}$ respectively,
which might be linked to the magneto-orthorhombic transition.

The evolution of $\chi_{0}^{B_{1g}}$ with temperature and pressure
is summarized in a colormap phase diagram in figure \ref{fig:chi0}(b).
Also plotted are the values of the structural ($T_{S}$), magnetic
($T_{m}$) and magneto-structural ($T_{S,m}$) transitions as reported
in \citep{Kothapalli2016a,Wang2016b}, along with the structural transition
temperatures extracted from our data as the temperature at which $\chi_{0}^{B_{1g}}$
is maximum. The latter are consistent with those of \citep{Kothapalli2016a,Wang2016b},
except at $\unit[1.6]{GPa}$ where our value appears somewhat higher.

The collapse of critical NF in the charge channel while magnetism
emerges upon increasing pressure  indicates that the nematic transition
is not magnetic-driven at low pressures, below $\sim\unit[1.6]{GPa}$,
but is rather driven by a $d$-wave Pomeranchuk instability of the
Fermi surface, whose associated charge NF contribute to Raman scattering
\cite{Gallais2016,Yamakawa2016,Chubukov2016a}. By contrast, at higher
pressures, the absence of critical NF can naturally be linked to a
fast weakening of this instability, and strongly suggests that the
orthorhombic distortion in this regime is only a mere consequence
of stripe-like magnetic ordering. The pressure induced change in the
driving force of nematicity naturally explains the absence of scaling
between the orthorhombic distortion and the ordered magnetic moment
reported recently by X-ray and Mössbauer measurements in FeSe under
pressure \cite{Boehmer2018}.

Our findings also imply a marginal role of critical NF in the enhancement
of $T_{c}$ observed between 2 and $\unit[6]{GPa}$, and their absence
at $\unit[5.8]{GPa}$ near the putative QCP contrasts with the divergent
nematic susceptibility observed near $T_{c,\mathrm{max}}$ in several
Fe SC \citep{Chu2012,Gallais2013,Kuo2016,Hosoi2016a}. Therefore,
while we cannot rule out a role of NF in the pairing mechanism, the
strong enhancement of $T_{c}$ in FeSe under pressure cannot be associated
to the presence of a nematic QCP \cite{Lederer2015,Labat2017}. It
is noteworthy that the disappearance of critical NF coincides with
the merging of the magnetic and structural transitions into a coupled
first-order transition, which occurs between $\unit[1.5]{GPa}$ and
$\unit[2.0]{GPa}$ (see figure \ref{fig:chi0}b) \citep{Kothapalli2016a,Wang2016b}.
As we show below, the evolution of optical phonons under pressure
points to a change in the low energy band structure in the same pressure
range, providing an underlying cause for these phenomena.

\begin{figure}
\begin{tikzpicture}
\node[anchor=south west,inner sep=0] (rp) at (0,0) {\includegraphics[width=1\columnwidth]{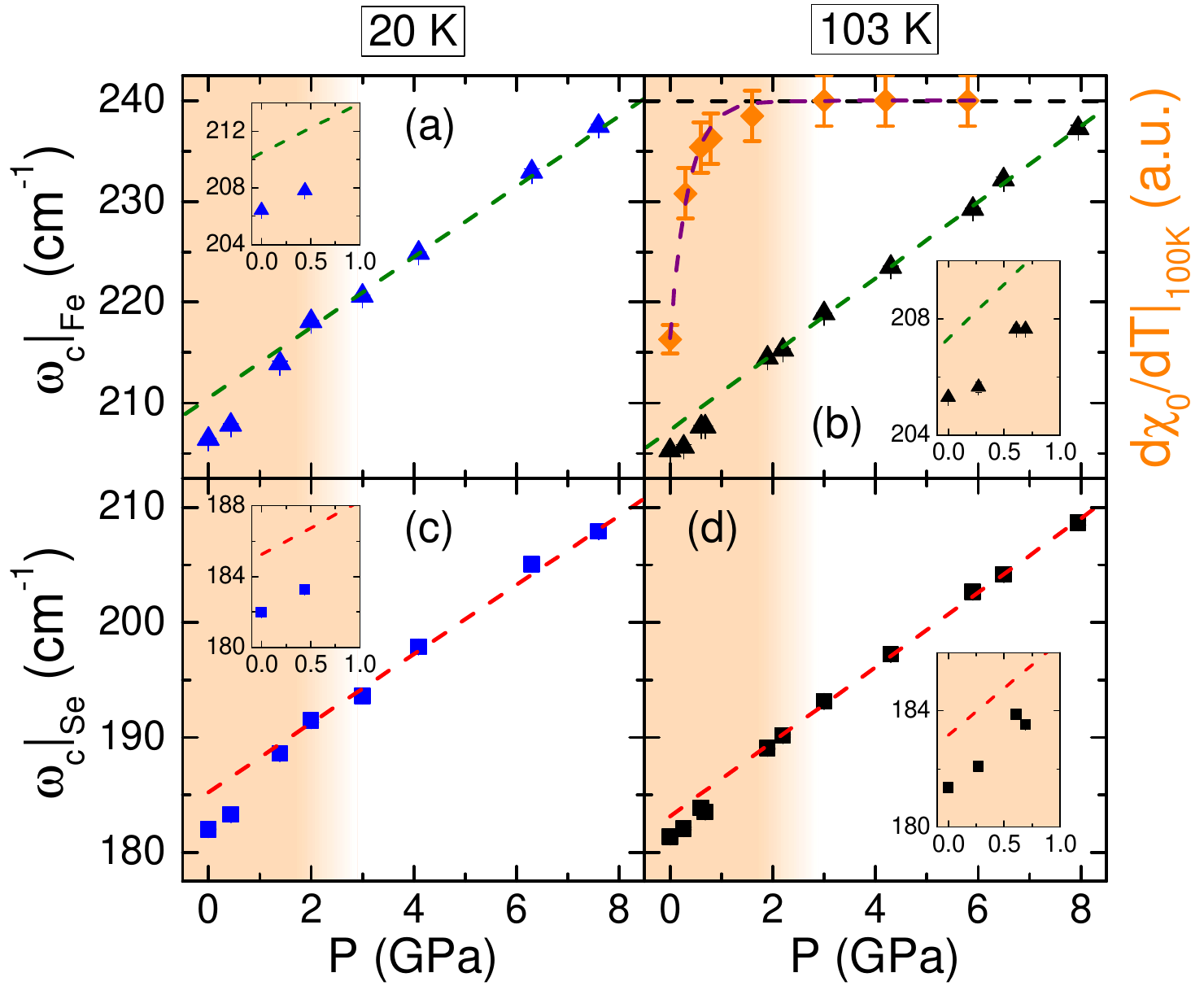}

};
\centering
\node[anchor=south west,inner sep=0] (b2g) at (3.23,3.59) {

\includegraphics[width=0.14\columnwidth]{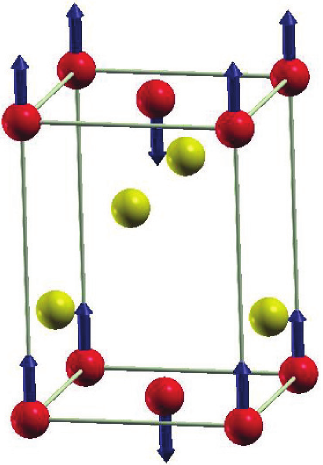}};
\centering
\node[anchor=south west,inner sep=0] (a1g) at (3.15,.9) {\includegraphics[width=0.15\columnwidth]{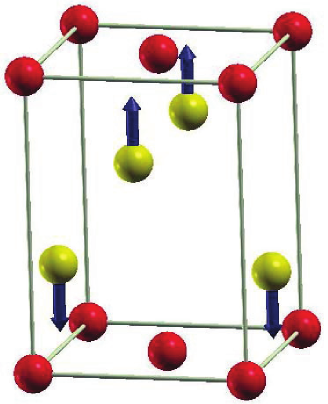}};	
\end{tikzpicture}

\caption{Pressure dependences of both Fe ($B_{2g}$, triangles, top row) and
Se ($A_{1g}$, squares, bottom row) phonons' energies at $\unit[20]{K}$
(left column, blue) and $\unit[103]{K}$ (right column, black). Dashed
lines are linear fits of the data points for $P>\unit[2]{GPa}$. Motions
of the atoms are shown as insets in (a) and (c): Fe and Se atoms are
in red and yellow respectively (reproduced from \cite{Ye2013a}).
An anomalous softening of $\unit[1-2]{\%}$ at $\unit[0]{GPa}$ is
clear from the low pressure zooms shown as insets. Also plotted in
(b) with orange diamonds is the pressure dependence of the slope of
the static susceptibility $\left.\frac{\mathrm{d}\chi_{0}^{B_{1g}}}{\mathrm{d}T}\right|_{\unit[100]{K}}$
as extracted from the linear fits in figure \ref{fig:chi0}(a); the
purple dashed line is a guide to the eye; the horizontal black dashed
line indicates $\mathrm{d}\chi_{0}/\mathrm{d}T=0$.\label{fig:phonons_positions}}
\end{figure}

In figure \ref{fig:phonons_positions} we plot the pressure evolutions
of the frequencies of two Raman active phonons at $\unit[103]{K}$
and $\unit[20]{K}$. The $A_{1g}$ (resp. $B_{2g}$) symmetry phonon
involves the motion of the Selenium (resp. Iron) atoms out of plane.
Between $\unit[2]{GPa}$ and $\unit[8]{GPa}$ the phonon frequencies
display a linear pressure hardening consistent with lattice contraction.
However below $\unit[2\pm0.5]{GPa}$ a clear frequency softening,
which manifests itself by a deviation from linearity in the pressure
dependence, is observed for both phonons. For the $B_{2g}$ Fe phonon
a linewidth broadening is also observed \citep{Massat2017SM}. Importantly,
while the observed deviations are stronger at $\unit[20]{K}$, they
are also visible at $\unit[53]{K}$ \citep{Massat2017SM} and $\unit[103]{K}$,
implying that they are not mere consequences of the magneto-structural
transition nor of the superconducting transition, which respectively
occur below $\unit[60]{K}$ and $\unit[40]{K}$ at all pressures \citep{Bendele2012,Miyoshi2013,Kothapalli2016a,Sun2016a}.
We note that anomalies of the structural parameters have been reported
in the same pressure range at low temperature \cite{Margadonna2009,Svitlyk2017}.

\begin{figure}[h]
\begin{tikzpicture}
\node[anchor=south west,inner sep=0] (phcalc) at (0,0) {\includegraphics[width=1\columnwidth]{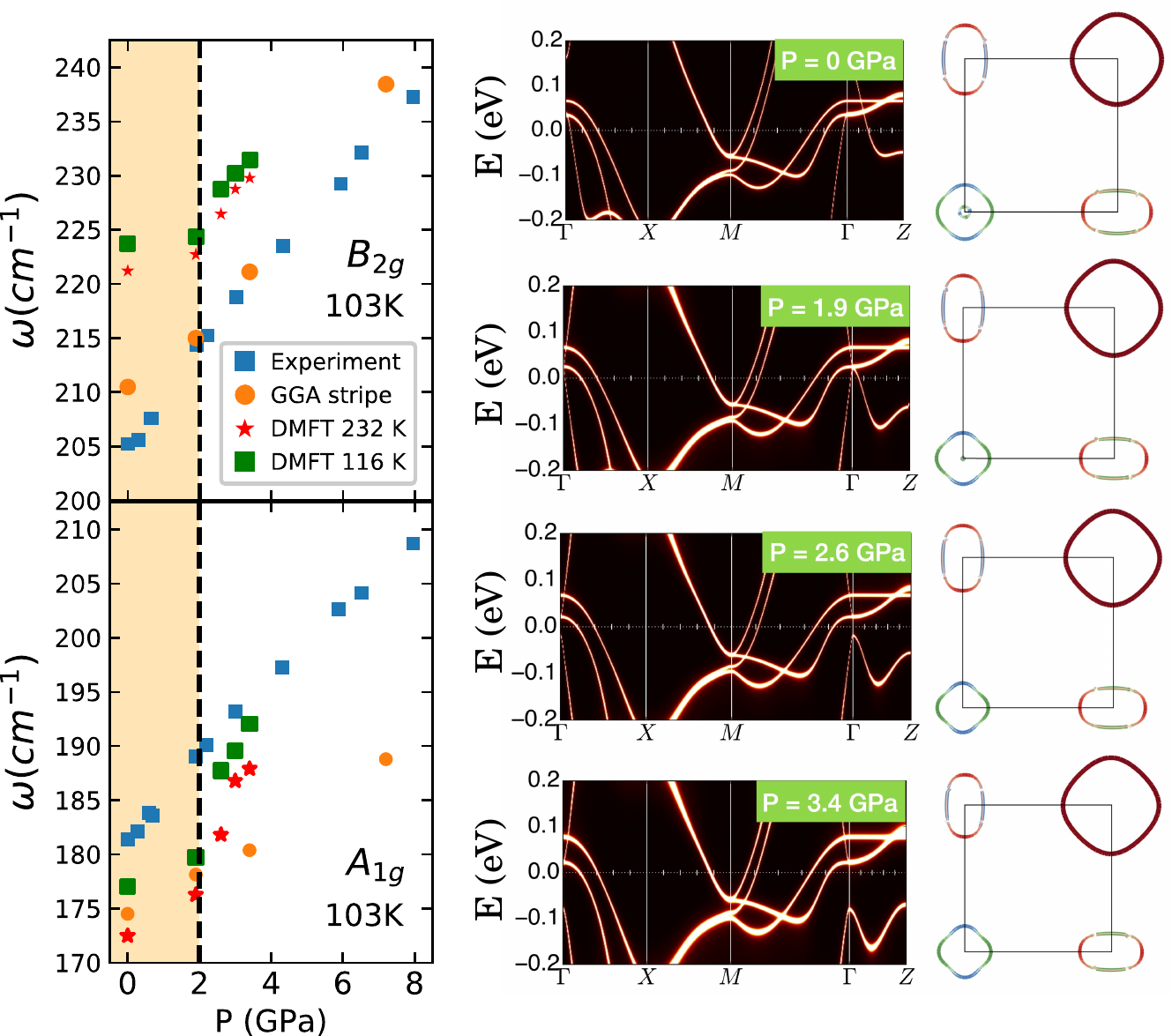}};
\begin{scope}[x={(phcalc.south east)},y={(phcalc.north west)}]
\node at (0.135,.92) {(a)};
\node at (0.135,.48) {(b)};
\node at (0.42,.97) {(c)};
\end{scope}
\end{tikzpicture}

\caption{LDA+DMFT calculations of FeSe phonons and band structure under pressure.
(a) $B_{2g}$ and (b) $A_{1g}$ phonon frequencies at $\boldsymbol{q}=\boldsymbol{0}$.
Blue squares: experimental data at $\unit[103]{K}$ (see figures \ref{fig:phonons_positions}).
Green squares and red stars: phonon frequencies obtained from LDA+DMFT
calculations in the PM state at $\unit[116]{K}$ and $\unit[232]{K}$
respectively. Orange circles: phonon frequencies from DFT–GGA calculations
in the stripe AFM state. (c) Band structure (left column) and orbitally-resolved
2D Fermi surface (right column) of FeSe calculated by LDA+DMFT at
$\unit[0]{GPa}$, $\unit[1.9]{GPa}$, $\unit[2.6]{GPa}$ and $\unit[3.4]{GPa}$.
Red, green and blue colors denote dominating $d_{xy}$, $d_{xz}$
and $d_{yz}$ orbital characters, respectively.\label{fig:LDA+DMFT-calculations}}
\end{figure}
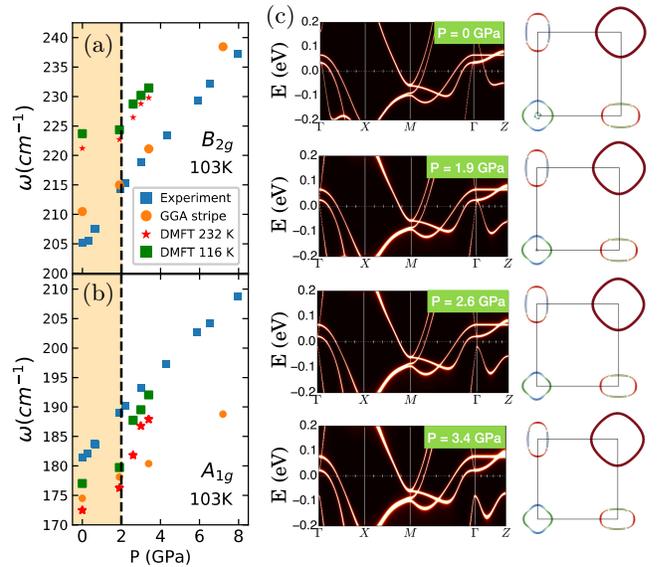

In order to clarify the origin of the phonon anomalies, we performed
theoretical calculations of the band structure and the phonon energies
under pressure, using local density approximation combined with dynamical
mean-field theory (LDA+DMFT, figure \ref{fig:LDA+DMFT-calculations}a,
b) \citep{Kotliar2006} in the paramagnetic state (PM). Remarkably
the calculations display a clear softening of both phonons' frequencies
at $\sim\unit[2]{GPa}$. In this pressure range, the calculated electronic
structure undergoes a Lifshitz transition due to the disappearance
of the inner hole pocket at the $\Gamma$ point between $\unit[1.9]{GPa}$
and $\unit[2.6]{GPa}$ (fig. \ref{fig:LDA+DMFT-calculations}(c),
see also \citep{Mandal2014}), suggesting the phonon anomalies are
associated to a change in Fermi surface topology \citep{Lifshitz1960}.
On the other hand, the phonon frequencies calculated using DFT in
the general gradient approximation (GGA) in the stripe magnetic phase,
where no Lifshitz transition occurs, do not show any significant anomaly
in the whole pressure range investigated (figure \ref{fig:LDA+DMFT-calculations}).
In this scenario the phonon softening below $\unit[2]{GPa}$ is associated
to the emergence of additional low energy electronic excitations to
which the optical phonons couple \cite{Cerdeira1972,Maksimov1996}.

The link between Fermi surface topology and the strength of critical
NF appears clearly when plotting together the evolution under pressure
of the Fe phonon frequency and that of the slope of the nematic susceptibility,
both at $\unit[103]{K}$ (figure \ref{fig:phonons_positions}b): the
latter goes to zero in the same pressure interval where the former
recovers linearity. This suggests that the low energy electronic states
responsible for the phonon softening at low pressure also drive strong
critical NF, causing a change in the regime of nematicity below $\sim\unit[2]{GPa}$.
Our findings are consistent with several theoretical studies which
have shown that the strength of the nematic and magnetic couplings
strongly depend on the size, the orbital content and the nesting conditions
of the hole and electron pockets \citep{Fernandes2012,Paul2014,Chubukov2016,Yamakawa2016,Fanfarillo2018, Classen2017, Ishizuka2018}.

To conclude it is interesting to contrast the evolution of critical
NF in FeSe under pressure with the case of 122 Fe SC. Recent works
have highlighted strong similarities between the pressure phase diagram
of FeSe and the doping phase diagram of BaFe$_{2}$As$_{2}$ \cite{Kothapalli2016a,Boehmer2018}.
However we note that critical NF are observed in both hole and electron
doped BaFe$_{2}$As$_{2}$ close to their optimal $T_{c}$ \cite{Boehmer2014,Kuo2016,Boehm2017}.
In addition, magnetic and nematic fluctuations are strongly linked
in BaFe$_{2}$As$_{2}$ \cite{Fernandes2013a} while they appear to
be essentially decoupled in FeSe. This suggests that while the phase
diagrams of both systems may appear similar from the point of view
of the ordered phases, they are fundamentally different in the nature
of the dominant fluctuations and their interplay.
\begin{acknowledgments}
P.M. and Y.G. are grateful to Lara Benfatto, Rafael Fernandes, Laura
Fanfarillo, Peter Hirschfeld, Alaska Subedi and Belen Valenzuela for
useful discussions. Y. G. also acknowledges Indranil Paul for numerous
insightful comments and discussions. P.T. acknowledges the financial
support of UGA and Grenoble INP through the AGIR-2013 contract of
S. Karlsson. Z.P.Y. and Y.Q. were supported by the National Natural
Science Foundation of China (Grant No. 11674030, 11704034), the Fundamental
Research Funds for the Central Universities (Grant No. 310421113)
and the National Key Research and Development Program of China through
Contract No. 2016YFA0302300. The calculations used high performance
clusters at the National Supercomputer Center in Guangzhou.
\end{acknowledgments}

The authors declare no competing interests.

%\bibliographystyle{apsrev4-1_nourl}
%\bibliography{../../../Biblio/biblio_FeSC,../../../Biblio/Techniques,additional_refs}


\end{document}